# FIRST RESULTS FROM Nb$_3$Sn COATINGS OF 2.6 GHz Nb SRF CAVITIES USING DC CYLINDRICAL MAGNETRON SPUTTERING SYSTEM*

M. S. Shakel[†], H. E. Elsayed-Ali, Old Dominion University, Norfolk, VA
G. Eremeev, Fermi National Accelerator Laboratory, Batavia, IL
U. Pudasaini, A. M. Valente-Feliciano, Thomas Jefferson National Accelerator Facility, Newport News, VA

*Abstract*

A DC cylindrical magnetron sputtering system has been commissioned and operated to deposit Nb$_3$Sn onto 2.6 GHz Nb SRF cavities. After optimizing the deposition conditions in a mock-up cavity, Nb-Sn films are deposited first on flat samples by multilayer sequential sputtering of Nb and Sn, and later annealed at 950 °C for 3 hours. X-ray diffraction of the films showed multiple peaks for the Nb$_3$Sn phase and Nb (substrate). No peaks from any Nb-Sn compound other than Nb$_3$Sn were detected. Later three 2.6 GHz Nb SRF cavities are coated with ~1 μm thick Nb$_3$Sn. The first Nb$_3$Sn coated cavity reached close to $E_{acc}$ = 8 MV/m, demonstrating a quality factor $Q_0$ of $3.2 \times 10^8$ at $T_{bath}$ = 4.4 K and $E_{acc}$ = 5 MV/m, about a factor of three higher than that of Nb at this temperature. $Q_0$ was close to $1.1 \times 10^9$, dominated by the residual resistance, at 2 K and $E_{acc}$ = 5 MV/m. The Nb$_3$Sn coated cavities demonstrated $T_c$ in the range of 17.9 – 18 K. Here we present the commissioning experience, system optimization, and the first results from the Nb$_3$Sn fabrication on flat samples and SRF cavities.

## INTRODUCTION

In comparison to bulk Nb SRF cavities, Nb cavities coated with Nb$_3$Sn demonstrate superior performance in terms of quality factors at 4 K and have the potential to replace bulk Nb cavities operated at 2 K [1-3]. Various processes have been applied for Nb$_3$Sn coating onto SRF cavities including Sn vapor diffusion, bronze routes, co-deposition of Nb and Sn by thermal evaporation, electrochemical synthesis, Nb dipping into Sn liquid followed by thermal diffusion, and magnetron sputtering [4-11]. Until now the vapor diffusion method has demonstrated the most promising radiofrequency (RF) outcomes for Nb$_3$Sn-coated cavities, whereas magnetron sputtering offers improved control over stoichiometry [12-13]. Magnetron sputtering can be employed in multiple techniques to grow Nb$_3$Sn, such as depositing multilayers of Nb-Sn followed by annealing, depositing from a single stoichiometric Nb$_3$Sn target, or co-sputtering of Nb and Sn [14-16].

Recently we commissioned a DC cylindrical magnetron sputtering system and developed methods for coating Nb$_3$Sn onto a 2.6 GHz Nb SRF cavity [17]. Our approach involves multilayer sequential sputtering using two identical cylindrical magnetrons to deposit Nb-Sn layers, which are subsequently annealed to promote the growth of Nb$_3$Sn. Initially, three Nb samples are positioned to mimic the beam tubes and the equator location of a 2.6 GHz Nb cavity. A deposition recipe is established to produce Nb$_3$Sn on the Nb samples and later the first Nb cavity (TTS1RI001) is coated following this recipe. Subsequently, optimization of the deposition conditions is carried out using a mock-up cavity and uniform thickness of Nb$_3$Sn is deposited on the beam tubes and the equator regions. Based on the refined conditions, two more 2.6 GHz Nb cavities (TTS1RI002 and TTS1RI003) are coated with Nb$_3$Sn following the multilayer sequential sputtering procedure.

## OPERATING CYLINDRICAL DC MAGNETRON SPUTTER COATER

The cylindrical magnetron sputtering setup comprises a custom-designed high-vacuum deposition chamber that houses a 2.6 GHz cavity with shape scaled from the TESLA center-cell shape and two identical cylindrical magnetrons [17]. The operating conditions of the magnetron discharge are optimized to achieve stable and symmetrical plasma formation for the Nb and Sn targets, installed on the top magnetron and the bottom magnetron, respectively, as shown in Fig. 1. With a baseline pressure of $\sim 2 \times 10^{-9}$ Torr, 30 W DC power is used for Nb discharge, and 8 W DC power is used for Sn discharge at 10 mTorr pressure with an Ar flow rate of 50 SCCM. To control the cylindrical sputter system and the associated devices, custom software was developed that is capable of running the operation of the deposition program and storing relevant data such as current, voltage, power, and water flow rate during the discharge process.

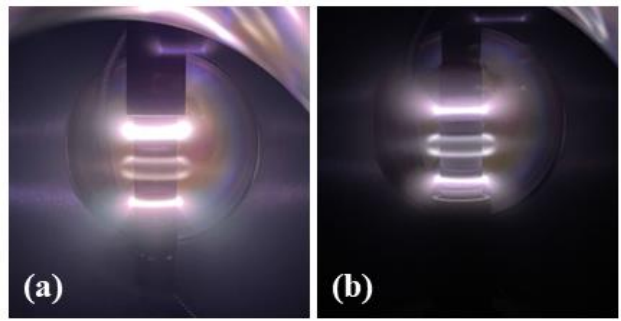

Figure 1: Photograph of stable plasma discharge at 10 mTorr using DC power (a) 30 W for Nb, (b) 8 W for Sn.

* Supported by DOE, Office of Accelerator R&D and Production, Contact No. DE-SC0022284, with partial support by DOE, Office of Nuclear Physics DE-AC05-06OR23177, Early Career Award to G. Eremeev.
† mshak001@odu.edu

# DEPOSITION OF Nb₃Sn ON Nb SAMPLES BY MULTILAYER SEQUENTIAL SPUTTERING METHOD

To optimize the deposition parameters for the Nb and Sn layers, we employed an aluminum mock-up cavity, shown in Fig. 2, designed to replicate the plasma geometry and density present within the 2.6 GHz Nb SRF cavity. Additionally, an aluminum sample holder is used to precisely position three Nb substrates (10 × 10 × 3 mm) at the inner surface locations of the cavity's beam tubes and the equator.

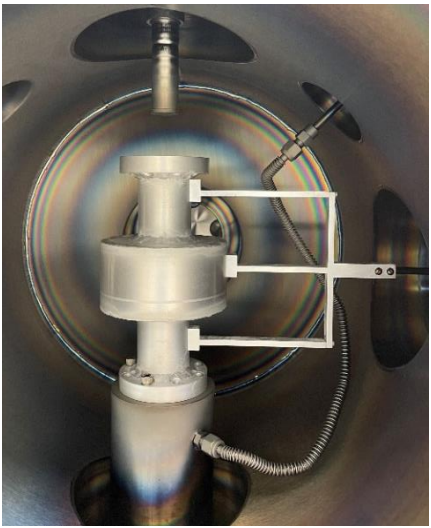

Figure 2: Al-made mock-up cavity and sample holder used for deposition conditions optimization.

To achieve a uniform thickness of Nb and Sn layers throughout the cavity, adjustments are made to the speed of magnetron movement and the number of scans. The top magnetron undergoes two passes across the top beam tube, equator, and bottom beam tube at speeds of 0.94 mm/s, 0.13 mm/s, and 1.02 mm/s, respectively, and deposits ~45 nm of Nb on three designated positions. The bottom magnetron makes a single pass across the top beam tube, equator, and bottom beam tube at speeds of 0.73, 0.20, and 1.10 mm/s, respectively, and deposits ~27 nm of Sn on three positions. By sequentially sputtering 14 layers of Nb and Sn, ~1 μm film is deposited on Nb samples mounted at three positions within the cavity. The thickness is determined through cross-sectional scanning electron microscopy. Subsequently, the samples undergo annealing at 950 °C for 3 hours in a separate furnace with a temperature ramp rate of 12 °C/min.

## RESULT AND DISCUSSION

The analysis of elemental compositions shows that the as-deposited samples obtained from the three designated positions exhibit Sn atomic compositions ranging from 32% to 35%. However, after the annealing process, the sample from the top beam tube position exhibits a Sn composition of approximately 22%, whereas the samples positioned at the equator and bottom beam tube positions demonstrate Sn compositions of approximately 12% and 14%, respectively. This observed reduction in Sn content during the annealing process aligns with previous findings in other studies involving the growth of Nb₃Sn using magnetron sputtering [14]. Figure 3 displays the X-ray diffraction (XRD) patterns for both the as-deposited and annealed samples. The XRD patterns of all three as-deposited samples exhibit distinct Nb diffraction peaks ((110), (200), (211), and (310)), originating from both the substrates and the deposited films. Additionally, the as-deposited sample obtained from the top beam tube position demonstrates several Sn diffraction peaks ((200), (101), (220), (211), (112), (400), and (312)).

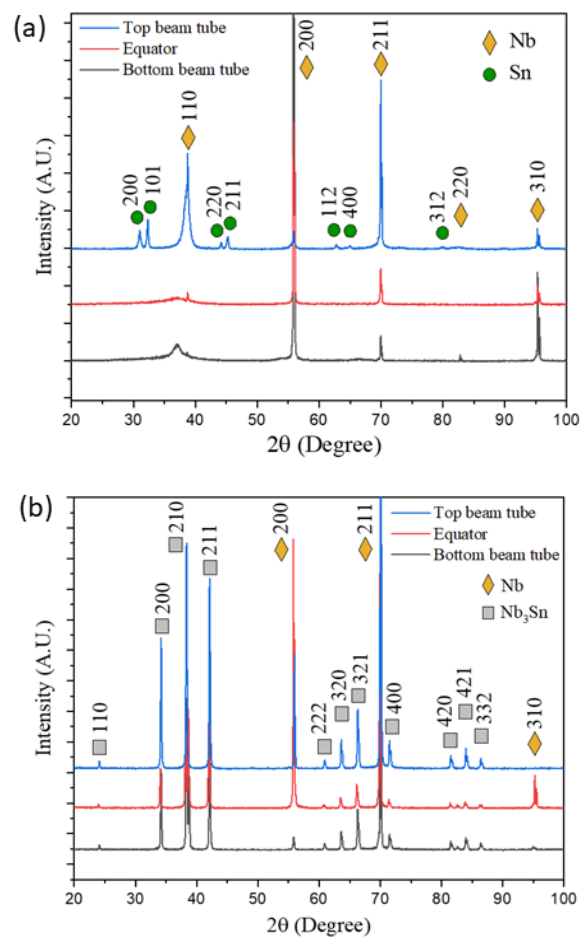

Figure 3: XRD patterns of (a) Nb-Sn as-deposited film, (b) annealed Nb₃Sn samples.

The X-ray diffraction (XRD) patterns of the annealed samples confirm the formation of Nb₃Sn. In the XRD patterns of the annealed samples from all three positions, multiple diffraction peaks corresponding to Nb₃Sn are observed, including (110), (200), (210), (211), (222), (320), (321), (400), (420), (421), and (332). No diffraction peaks indicative of Nb-Sn intermetallic phases other than Nb₃Sn are detected, providing conclusive evidence that the thin film consists of Nb₃Sn only. Additionally, the XRD patterns of the annealed films reveal the presence of several Nb peaks ((200), (211), (310)), originating from the substrate.

## Nb$_3$Sn COATING ONTO 2.6 GHz Nb CAVITIES

With the optimized deposition conditions in the mock-up cavity, two 2.6 GHz Nb cavities (TTS1RI002 and TTS1RI003) are coated with ~1 μm thick Nb$_3$Sn film throughout the cavity following the multilayer sequential sputtering of Nb and Sn layers. Note that, the first coated cavity TTS1RI001 was coated with ~1.2 μm Nb$_3$Sn on the equator and ~880 nm and ~816 nm on the top beam tube and the bottom beam tube respectively [17]. Following the sputtering process, the coatings within the cavities exhibited a uniform appearance, without any visible film peeling.

Next, the coated cavities were annealed in a separate high-vacuum furnace. This furnace is comprised of a hot zone measuring 12" x 12" x 18" enclosed with molybdenum heat shields. Prior to annealing, customized molybdenum covers were assembled onto the cavity flanges. The furnace containing the cavity setup was evacuated to a range of 10$^{-7}$ Torr. The hot zone, along with the cavity, underwent a degassing step at approximately 300 °C for a duration of 24 hours. The primary annealing process was conducted at 950 °C for a period of 3 hours. Following annealing, a variation in color appearance of the film was observed within the cavity, as shown in Fig. 4.

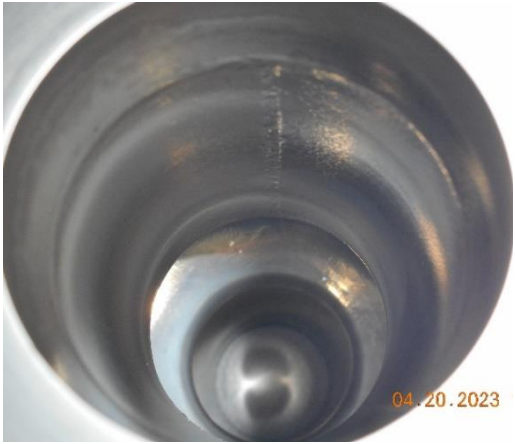

Figure 4: Pictures of post-annealed picture of Nb$_3$Sn coated 2.6 GHz Nb cavity (TTS1RI002).

Following the annealing process, the cavity is prepared for cryogenic RF testing in the cleanroom where it received high-pressure water rinsing using ultra-high purity water. Two cavities, TTS1RI001 and TTS1RI002, have been tested. During the cryogenic RF test of the first coated cavity (TTS1RI001), it demonstrated a quality factor $Q_0$ of 3.2 x 10$^8$ at $E_{acc}$ = 5 MV/m at $T_{bath}$ = 4.4 K and $Q_0$ of 1.1 x 10$^9$ at $E_{acc}$ = 5 MV/m at $T_{bath}$ = 2 K, as shown in Fig. 5. The second coated (TTS1RI002) cavity demonstrated $Q_0$ of 9.3 x 10$^7$ at $E_{acc}$ = 1 MV/m at $T_{bath}$ = 4.4 K and $Q_0$ of about 1 x 10$^9$ at $E_{acc}$ = 1 MV/m at $T_{bath}$ = 2 K. The second cavity was limited by a strong Q-slope. Note that, the bulk 2.6 GHz Nb cavity has a $Q_0$ of about 1 x 10$^8$ at $T_{bath}$ = 4 K and $Q_0$ of about 7 x 10$^9$ at $T_{bath}$ = 2 K.

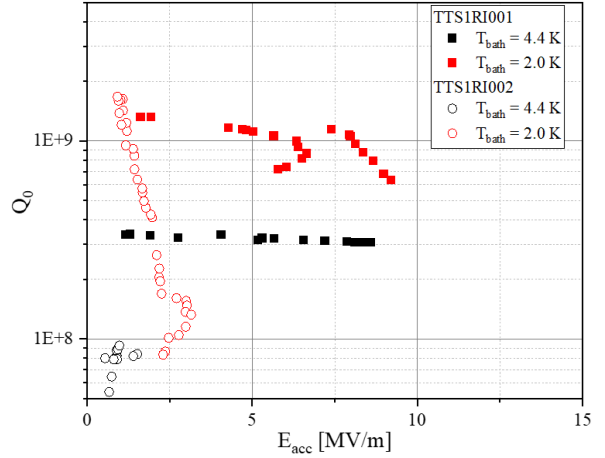

Figure 5: $Q_0$ versus $E_{acc}$ at 4.4 K and 2.0 K for the two Nb$_3$Sn-coated 2.6 GHz Nb cavities.

Using a network analyzer, the $Q_L$ of the coated cavities was measured as a function of temperature from the width of the resonant peak in S21 transmission of one of the resonant peaks using FWHM technique. As demonstrated in Fig. 6, in the case of TTS1RI002, the superconducting transition is seen close to 18 K, as expected for Nb$_3$Sn. Another superconducting transition is observed at below 9 K. This transition may be caused by Sn-deficient regions close to the ends of coated cavity. In the case of TTS1RI001 cavity, one superconducting transition was measured close to 17.9 K, another transition was seen close to 9.2 K, speculated to be from Nb flanges, and another superconducting transition was observed below 9 K.

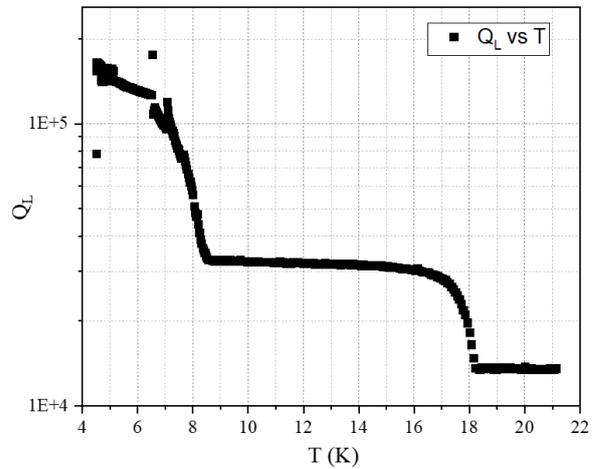

Figure 6: $Q_L$ versus temperature (T) of Nb$_3$Sn coated cavity (TTS1RI002) showing the superconducting transition temperature at about 18 K due Nb$_3$Sn.

## SUMMARY

Operating the recently commissioned DC cylindrical magnetron sputtering system, we have optimized deposition conditions in the mock-up cavity and followed multilayer sequential sputtering technique to fabricate ~1 µm thick $Nb_3Sn$ films on Nb samples at the beam tubes and the equator positions. Following the same coating recipe, we coated three 2.6 GHz Nb SRF cavities and later annealed them at 950 °C for 3 hours. RF testing of the two cavities demonstrated superconducting transition temperature between 17.9 – 18 K. One of the cavities demonstrated (TTS1RI001) $Q_0$ of $3.2 \times 10^8$ at $E_{acc}$ = 5 MV/m at $T_{bath}$ = 4.4 K and $Q_0$ of $1.1 \times 10^9$ at $E_{acc}$ = 5 MV/m at $T_{bath}$ = 2 K. The other cavity (TTS1RI002) demonstrated $Q_0$ of $9.3 \times 10^7$ at $E_{acc}$ = 1 MV/m at $T_{bath}$ = 4.4 K and $Q_0$ of about $1 \times 10^9$ at $E_{acc}$ = 1 MV/m at $T_{bath}$ = 2 K.